\documentclass[10pt,conference]{IEEEtran}
\usepackage[utf8]{inputenc}
\usepackage{cite}
\usepackage{url}
\usepackage[pdftex]{graphicx}
\graphicspath{{./fig/}}
\DeclareGraphicsExtensions{.pdf,.jpeg,.png}
\usepackage[cmex10]{amsmath}
\usepackage{mathtools}
\usepackage{algorithmic}
\usepackage{array}
\usepackage{mdwmath}
\usepackage{mdwtab}
\usepackage{eqparbox}
\usepackage{fixltx2e}
\usepackage[caption=true,font=footnotesize]{subfig}
\usepackage{pgfplots}
\begin{document}

\title{Detour Planning for Fast and Reliable \\ Failure Recovery in SDN with OpenState}

\author{
  \IEEEauthorblockN{
    Antonio Capone\IEEEauthorrefmark{1},
    Carmelo Cascone\IEEEauthorrefmark{1}\IEEEauthorrefmark{2},
    Alessandro Q.T. Nguyen\IEEEauthorrefmark{1}, 
    Brunilde Sansò\IEEEauthorrefmark{2}
  }
  \IEEEauthorblockA{
    \IEEEauthorrefmark{1}
    Dipartimento di Elettronica, Informazione e Bioingegneria, Politecnico di Milano, Italy\\
    Email: antonio.capone@polimi.it,
    alessandro.nguyen@mail.polimi.it
  }
  \IEEEauthorblockA{
    \IEEEauthorrefmark{2}
    D\'epartement de g\'enie \'electrique, \'Ecole Polytechnique de Montr\'eal, Canada\\
    Email: carmelo.cascone@polymtl.ca,
    brunilde.sanso@polymtl.ca
  }
}


\maketitle

\begin{abstract}
A reliable and scalable mechanism to provide protection against a link or node failure has additional requirements in the context of SDN and OpenFlow. Not only it has to minimize the load on the controller, but it must be able to react even when the controller is unreachable. In this paper we present a protection scheme based on precomputed backup paths and inspired by MPLS ``crankback'' routing, that guarantees instantaneous recovery times and aims at zero packet-loss after failure detection, regardless of controller reachability, even when OpenFlow's ``fast-failover'' feature cannot be used. The proposed mechanism is based on OpenState, an OpenFlow extension that allows a programmer to specify how forwarding rules should autonomously adapt in a stateful fashion, reducing the need to rely on remote controllers. We present the scheme as well as two different formulations for the computation of backup paths.
\end{abstract}

\section{Introduction}
\label{sec:introduction}

Failure management is one of the fundamental instruments that allows network operators to provide communication services that are much more reliable than the individual network components (nodes and links). It allows reacting to failures of network components by reconfiguring the resource allocation so as to make use of the surviving network infrastructure able to provide services.

Traditionally, failure resilience has been incorporated in distributed protocols at the transport (like e.g. SDH) and/or network layer (like e.g. MPLS) with some optimization of resources pre-computed for a class of possible failures (like e.g. single link or node failures) and implemented with signaling mechanisms used to notify failures and activate backup resources.

With the introduction of the revolutionary and successful paradigm of Software-defined Networking (SDN), the traditional distributed networking approach is replaced with a centralized network controller able to orchestrate traffic management through the programing of low-level forwarding policies into network nodes (switches) according to simple abstractions of the switching function like that defined in OpenFlow with the match/action flow table \cite{mckeown08}.

Even if SDN and OpenFlow provide huge flexibility and a powerful platform for programming any type of innovative network application without the strong constraints of distributed protocols, they can make the implementation of important traditional functions, like failure resilience, neither easy nor efficient, since reaction to events in the network must always involve the central controller (notification of an event and installation of new forwarding rules) with non-negligible delays and signaling overheads.

New versions of OpenFlow \cite{of14} have recently introduced a mechanism, namely fast-failover, for allowing quick and local reaction to failures without the need to resort on central controller. This is obtained by instantiating multiple action buckets for the same flow entry, and applying them according to the status of links (active or failed). However, fast-failover can be used only to define local detour mechanisms when alternative paths are available from the node that detects the failure. Depending on network topology and the specific failure, local detour paths may not be available or they may be inefficient from the resource allocation point of view.

A recent proposal (by some of the authors) \cite{bianchi14,bianchi14b}, named OpenState, has extended the data plane abstraction of OpenFlow to include the possibility for switches to apply different match-actions rules depending on states and to make states evolve according to state machines where transitions are triggered by packet-level events.

In this paper, we propose a new approach to failure management in SDN which exploits OpenState ability to react to packet-level events in order to define a fast path restoration mechanism that allows to reallocate flows affected by failure by enabling detours in any convenient nodes along the primary path. No specific signaling procedure is adopted for triggering detours, rather the same packets of the data traffic flows are tagged and forwarded back to notify nodes of the failure and to induce a state transition for the activation of pre-computed detours.

We define optimization models for the computation of backup paths for all possible single node and link failures that consider multiple objectives including link congestion level, distance of the reroute point from the failure detection point, and level of sharing of backup paths by different flows. We show that the MILP (Mixed Integer Linear Programming) formulations proposed are flexible enough to incorporate the optimization of the OpenFlow fast-failover reroutes as a special case and that path computation for all possible failure scenarios can be performed within reasonable time for realistic size networks with state-of-the-art solvers (cplex).

The reminder of the paper is presented as follows. In Section~\ref{sec:openstate} we first present an overview of OpenState and next we  present the proposed failure recovery scheme in Section~\ref{sec:approach}. Related work is reviewed in Section~\ref{sec:related} and in Section~\ref{sec:model} two modelling formulations are presented. Computational results are discussed in Section~\ref{sec:results}. Conclusions are provided in Section~\ref{sec:conclusion}.

\section{OpenState}
\label{sec:openstate}

The most prominent instance of SDN is OpenFlow, which, by design, focuses on an extreme centralization of the network intelligence at the controller governing switches, which in turn are considered dumb. In OpenFlow, adaptation and reconfiguration of forwarding policies can only be performed by remote controllers, with a clear consequence in terms of overhead and control latency. OpenState is an OpenFlow extension that enables mechanisms for controllers to offload some of their control logic to switches. In OpenState, the programmer is able to define forwarding rules that can autonomously adapt in a stateful fashion on the basis of packet-level events. The motivation beside OpenState is that control tasks that require only switch-local knowledge are unnecessarily handled at the controller, and thus can be offloaded to switches, while maintaining centralized control for those tasks that require global knowledge of the network.

\begin{figure}
  \centering
  \includegraphics[width=\columnwidth]{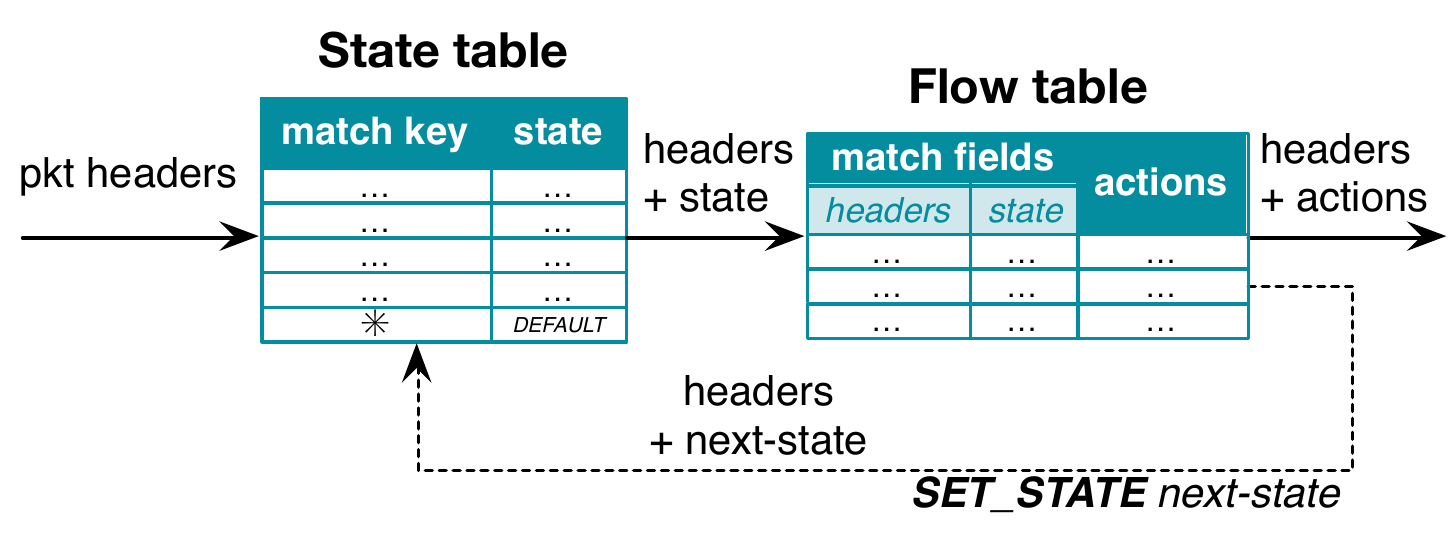}
  \caption{Simplified packet flow in OpenState.}
  \label{fig:os-stateful-stage}
  \vspace{-5mm}
\end{figure}

OpenState has been designed as an extension (superset) of OpenFlow. In OpenState the usual OpenFlow match/action flow table is preceded by a state table that contains the so called ``flow-states''. First, packets are matched against the state table using only a portion of the packet header (a programmable lookup-key), a state lookup operation is performed and a state label (similar to OpenFlow's metadata) is appended to packet headers. A \texttt{DEFAULT} state is returned if no row is matched in the state table. Packets are then sent to the flow-table where the usual OpenFlow processing is performed, while a new \texttt{SET\_STATE} action is available to insert or rewrite rows of the state table with arbitrary values. Figure~\ref{fig:os-stateful-stage} illustrates the packet flow in the two tables. OpenState allows also to match packets using ``global-states'', so called because, in contrast to flow-states, these are globally valid for the whole switch (datapath) and not just for a given flow. By using flow-states and global-states a programmer can define flow entries that apply to different scenarios, and by using state transition primitives she can control how those scenarios should evolve.

OpenState has been showed to bring tangible benefits in the implementation of fundamental network applications \cite{bianchi14b}. An open-source implementation of an OpenState controller and switch can be found at \cite{openstatehomepage}, along with a modified version of Mininet and few application examples.

\section{Proposed Approach}
\label{sec:approach}

The approach we take is similar to that used in crankback signaling \cite{rfc4920}. In the context of end-to-end QoS in MPLS and GMPLS with RVSP-TE, when a connection or flow setup fails because of a blocked link or node, crankback is a signaling technique in which a notification of the failure is backtracked along the flow path, from the upstream node that faces the blockage up to the first node (called ``repair point'') that can determine an alternative path to the destination node.

\begin{figure}
  \centering
  \includegraphics[width=\columnwidth]{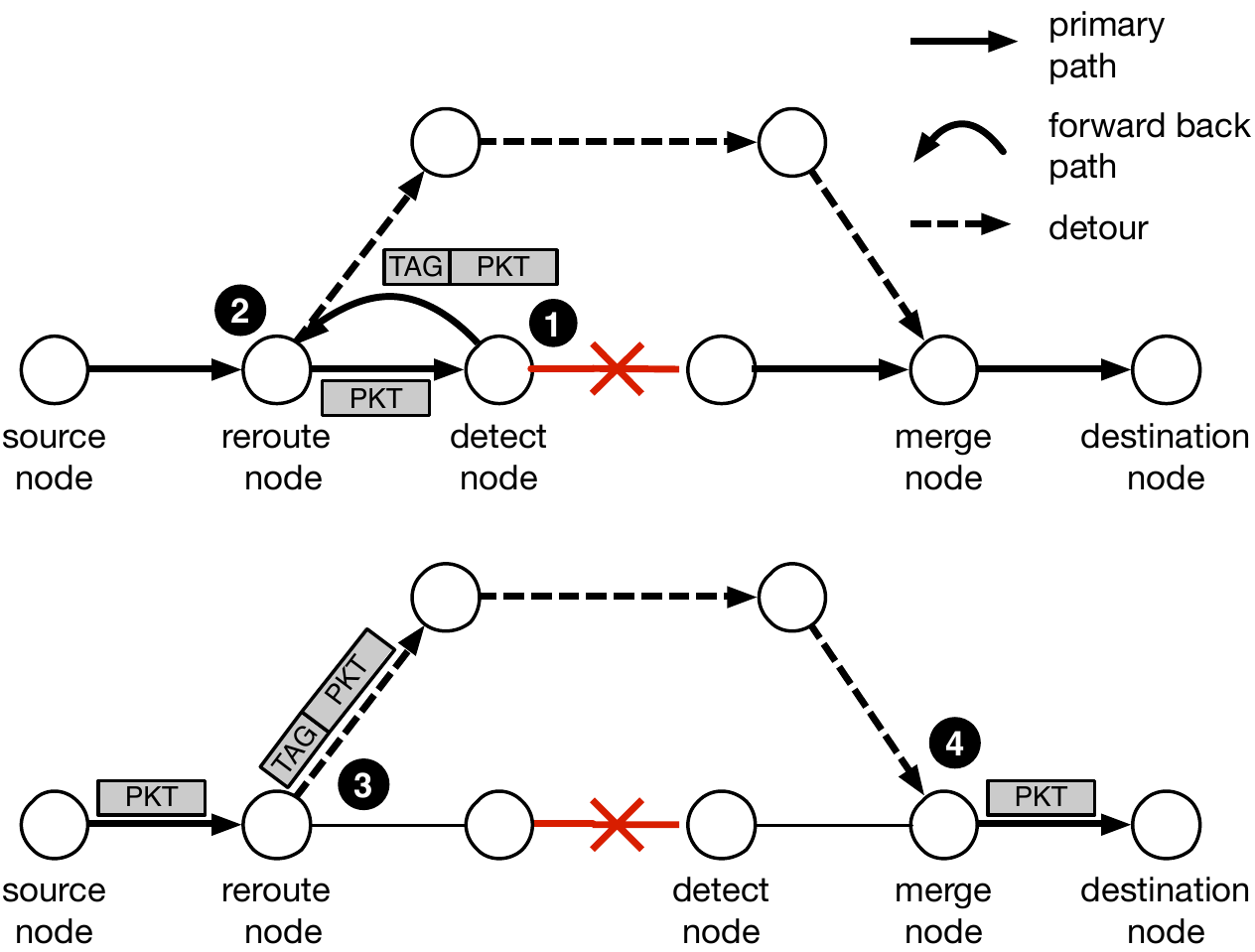}
  \caption{Example of failure recovery with OpenState: in (1) the upstream node detects the failure, tags the packet and forwards it back. In (2) the reroute node receive the tagged packet, executes a state transition and forward the packet on the detour. In (3) all the packets received for the considered demand after the state transition, will be tagged and forwarded on the detour. Finally in (4), at the end of the detour, the tag is popped and the packet is forwarded on the primary path, towards its destination node.}
  \label{fig:os-ft-example}
  \vspace{-5mm}
\end{figure}

Our solution is based on the same idea, but with the major difference that, upon link or node failure, the same data packets, and not a notification, can be sent back on their original path. We distinguish two situations: (i) the node which detects the failure is able to reroute the demand, and (ii) the packet must be forwarded back on it's primary path until a convenient reroute node is encountered. In the first case, solutions like OpenFlow's fast-failover already guarantee almost instantaneous protection switching without controller intervention, while in the second case it would be impracticable to signal the failure to other nodes without the intervention of the controller. The novelty of our approach is given by the fact that, in the second case, a crankback approach is performed using the same data packets, which are first tagged (e.g. with a MPLS label containing information on the failure event) and then sent back trough the primary path. A reroute node who receives the tagged packet will be able to respond to the failure by rerouting the tagged packets and by enabling a detour for all subsequent packets. That said, only the first packets of the flow are sent back from the detect node. As soon as the first tagged packet is processed by the reroute node, a state transition is performed in the OpenState switch, and all subsequent packets coming from the source node will be forwarded on the detour. An example of the mechanism described so far is summarized in Figure~\ref{fig:os-ft-example}.

With this approach, flow-states are used to distinguish the forwarding of each traffic demand at each switch. The \texttt{DEFAULT} state implies that the demand can be forwarded towards the next node on the primary path, other arbitrary states are used to describe the specific failure that can affect the demand, so that the same reroute nodes can react differently according to the specific failure event. Global-states are instead used to describe the operational status of switch ports (up or down). In this case global-states are completely equivalent to ``port liveness'' states used by OpenFlow fast-failover feature.

Our proposal is currently independent of the way failures are detected, because it does not influence the modeling aspect of the solution. We assume it could be implemented either via Loss Of Signal (LOS) or Bidirectional Forwarding Detection (BFD) \cite{rfc5880} mechanisms. In both cases, as soon as the state of the failed port is updated, our solution guarantees instantaneous reaction with ideally zero packet-loss.

\section{Related Work}
\label{sec:related}

Failure management in SDN is a topic that has been already explored by the research community. In \cite{staessens11} the authors analyze the case of restoration for OpenFlow networks, showing how hard it is to achieve fast ($<50$ms) recovery times in large networks. Restoration is also taken into consideration in \cite{sharma11}, where the controller is entitled to monitor link status on the network, and, in case of failure, it computes a new path for the affected demand and replaces or deletes flow entries in switches, accordingly. In \cite{kempf12} an end-to-end path protection scheme is proposed: OpenFlow 1.1 is extended by implementing in the switches a monitoring function that allows to reduce processing load on the controller. Such a function is used in conjunction with OpenFlow fast-failover feature, thus allowing nodes to autonomously react to failures by switching to a precomputed end-to-end backup path. In \cite{sgambelluri13} a segment protection mechanisms is proposed  only for the case of link failure. Backup paths are pre-installed, and OpenFlow is extended to enable switches to locally react to connected failed links. Another way to reduce 
the load at the controller is presented in \cite{lee14}. The authors propose a centralized monitoring scheme and a model to reduce the number of monitoring iterations that the controller must perform in order to check all links. A completely different and creative approach is proposed in \cite{borokhovich14}, where classic graph search algorithms are presented to implement a solution based on the OpenFlow fast-failover scheme, where backup paths are not known in advance but nodes implement an algorithm to randomly try new routes to reach the destination.

\section{Problem Formulation}
\label{sec:model}


Let $G(N,A)$ be a symmetric directed graph so that $N$ represents the set of network switches, and $A$ the set of links between switches. The demands  are assumed to be known in advance. Also assumed is the fact that  each demand will be routed using a primary path optimized as a shortest-path with link capacity constraints. Our main problem then focuses on the evaluation of backup paths for each demand, for every possible single failure scenario in the primary path. 
The significance of a failure scenario will be clearly indicated in the next subsection. For comparison purposes we also present, at the end of the Section, a congestion avoidance version of the same backup path problem.

\subsection{Backup Path Problem Formulation}
\label{sec:bp-problem}

In the forthcoming model, we refer to  ``failure detection event" rather than simply ``failure state" to indicate that a failure has been perceived. Moreover,
instead of making an a priori distinction between the case of link and the case of node failure,   a ``fault detection event'' $f = (n,m)$ may be either. The notation simply indicates that node $n$ detects a failure while transmitting to a downstream node $m$. Therefore two distinct situations are considered: (i) a failure on link $(n,m)$ (e.g. disconnected or truncated cable, etc.) and (ii) a scenario where downstream node $m$ fails implying the disconnection of all its adjacent links. When evaluating the backup path for a given demand, we always consider the worst-case scenario of a node failure, thus completely avoiding to forward packets to $m$, except for the case where $m$ is also the destination node of the considered demand ($m = t_d$). In such a case, we try to deliver packets to $m$ avoiding the failed link $(n,m)$. 

Let us now define the following parameters:

\subsection*{Parameters}

\begin{description}[\IEEEsetlabelwidth{$u_{ij}^{nm}$}]
  \item[D] set of demands to be routed;
  \item[$s_d$] source node of demand $d$;
  \item[$t_d$] destination node of demand $d$;
  \item[$\beta_{dij}$] is equal to 0 if link $(i,j)$ belongs to the primary path for demand $d$, otherwise 1;
  \item[$b_d$] requested bandwidth for demand $d$;
  \item[$c_{ij}$] total capacity of link $(i,j)$;
  \item[$w_{cap}$] percentage of the link capacity available;
  \item[$F$] set including all the possible failure detection events $(n,m)$ that can affect at least one primary path;
  \item[$D^{nm}$] subset of D including all the demands affected by the failure detection events $(n,m)$;
  \item[$D_1^{nm}$] subset of $D^{nm}$ including all the demands $d$ affected by the failure detection event $(n,m)$, when $m$ is not the destination node of the considered demand and thus $m \neq t_d$;
  \item[$D_2^{nm}$] subset of $D^{nm}$ including all the demands $d$ affected by the failure detection event $(n,m)$, where $m$ is the destination node of the considered demand and thus $m=t_d$;
  \item[$L^{m}$] subset of A that will include all the links incident to node $m$;
  \item [$u_{ij}^{nm}$] represents the used capacity of link $(i,j)$ when link $(n,m)$ fails. Note that in this parameter we consider only the link capacity allocated for those demands for which the primary path does not include neither $(m,n)$ or $(n,m)$;
  \item [$v_{ij}^{m}$] is the used capacity of link $(i,j)$ in case of failure for node $m$. In this case we consider only the link capacity allocated for those demands that are not affected by a failure of node $m$, in other words those demands which primary path does not include any of the links incident to $m$;
  \item [$p_d^k$] represents the link $(i,j)$ in the $k$-th position of the primary path for demand $d$, where $k=1$ is intended as the first link of the primary path starting from node $s_d$;
  \item [$\lambda^{nm}_d$] is the number of nodes that a packet of demand $d$ traverses on the primary path, before reaching node $n$ of failure detection event $(n,m)$. $\lambda^{nm}_d = 0$ means that the failure has been detected by the first node of the path.
\end{description}

\subsection*{Decision variables}

\begin{description}[\IEEEsetlabelwidth{$y_{dij}^{nm})$}]
  \item[$y_{dij}^{nm}$] is equal to 1 if link $(i,j)$ belongs to the backup path of demand $d$ in case of failure detection event $(n,m)$, otherwise 0;
  \item[$h^{nm}_d$] non-negative integer that represents the number of backward hops that a tagged packet of demand $d$ must perform in case of failure detection event $(n,m)$, before reaching the reroute node that will enable the detour. When $h^{nm}_d = 0$ we mean that node $n$ that detected the failure is also the reroute node;
  \item[$z_{dij}$] equal to 0 if $(i,j)$ is not used by any backup path (for every possible failure) for demand $d$, otherwise 1.
\end{description}

\subsection*{Objective Function}
\vspace{-5mm}

\begin{IEEEeqnarray}{lCl}
  min &   & \sum_{(n,m) \in F}\sum_{d \in D^{nm}} w_{h}h^{nm}_d \nonumber \\
      & + & \sum_{(n,m) \in F} \sum_{d \in D^{nm}}\sum_{(i,j) \in A} w_{y}y^{nm}_{dij} \nonumber \\
      & + & \sum_{d \in D} \sum_{(i,j) \in A}w_{z} \beta_{dij} z_{dij}
  \label{eq:bp-obj}
\end{IEEEeqnarray}
The objective function is composed of three weighted terms. The first minimizes  the length of the reverse path that tagged data packets must travel in case of failure. The second minimizes the length of backup paths. The third term minimizes the number of links allocated to the backup paths for a given demand, in other words we want more backup paths of the same demand to share the same links. By using the three weights $w_h$, $w_y$, and $w_z$ we are able to characterize the behavior of the objective function in different ways.

\subsection*{Link availability constraints}
\vspace{-5mm}

\begin{IEEEeqnarray}{c}
  \sum_{(i,j) \in L^m} y^{nm}_{dij}\le 0
  \ \ \ \
  \forall (n,m) \in F, \forall d \in D_1^{nm}
  \label{eq:bp-la-cons-generic} 
\end{IEEEeqnarray}
\begin{IEEEeqnarray}{c}
  y^{nm}_{dnm} + y^{nm}_{dmn}\le 0
  \ \ \ \
  \forall (n,m) \in F, \forall d \in D_2^{nm}
  \label{eq:bp-la-cons-terminal} 
\end{IEEEeqnarray}

These constraints disable the use of certain links when evaluating the backup path for a given demand.

\subsection*{Link capacity constraints}
\vspace{-5mm}

\begin{IEEEeqnarray}{c}
  u^{nm}_{ij}
  + \sum_{d \in D^{nm}} b_d y^{nm}_{dij}
  + \sum_{e \in D^{mn}} b_e y^{mn}_{eij}
  \le w_{cap} c_{ij} \nonumber \\
   \forall (n,m) \in F, \forall (i,j) \in L
  \label{eq:bp-lc-cons-link}
\end{IEEEeqnarray}

\begin{IEEEeqnarray}{c}
  v^{m}_{ij}
  + \sum_{\substack{n \in N: \\ (n,m) \in F}} \sum_{d \in D^{nm}} b_d y^{nm}_{dij}
  \le  w_{cap}c_{ij} \nonumber\\
  \forall m \in N, \forall (i,j) \in L
  \label{eq:bp-lc-cons-node}
\end{IEEEeqnarray}

The above constraints insure that for every possible failure, when allocating the backup paths, the link capacity must be respected. The first set of constraints is specific for the case of link failure, while the second set is specific for the case of node failure. Because we do not know the exact nature of a failure detection event, we want our solution to be valid (in terms of resource allocation) in case of both link and node failure.

\subsection*{Flow conservation constraints}
\vspace{-5mm}

\begin{IEEEeqnarray}{c}
  \sum_{\mathclap{\substack{
          j \in N: \\
          (i,j)\in A}}}
      y^{nm}_{dij}
  - \sum_{\mathclap{\substack{
      j \in N: \\
      (j,i) \in A}}}
    y^{nm}_{dji}
  =
  \left\{
    \begin{array}{ll}
    1,  & \text{if} \ i =  s_d; \\ 
    -1, & \text{if} \ i = t_d ; \\
    0,  & \text{otherwise.}
    \end{array}
  \right. \nonumber \\
  \forall i \in N, \forall (n,m) \in F, \forall d \in  D^{nm}
  \label{eq:bp-fc-cons} 
\end{IEEEeqnarray}

These constraints assure that there is continuity in backup paths. 

\subsection*{Cycle avoidance constraints}
\vspace{-5mm}

\begin{equation}
  \sum_{\mathclap{\substack{
    j \in N: \\
    (i,j)\in L}}} 
  y^{nm}_{dij}
  \le 1
  \ \ \ \
  \forall i \in N, \forall (n,m) \in F, \forall d \in  D^{nm}
  \label{eq:nocycle}
\end{equation}

These constraints avoid the creation of cycles in the backup paths.
 
\subsection*{Reverse path constraints}
\vspace{-5mm}

\begin{IEEEeqnarray}{c}
  \sum_{\substack{k=1:\\ (i,j) = p_d^k}}^{\lambda^{nm}_d}
    (1-y_{dij}^{nm}) \le h^{nm}_d
  \nonumber \\
  \forall (n,m) \in F, \forall d \in D^{nm} : \lambda^{nm}_d\neq 0
   \label{eq:reverseBP}
\end{IEEEeqnarray}

These constraints are needed to evaluate the variable $h^{nm}_d$.

\subsection*{Capacity use constraints}
\vspace{-5mm}

\begin{IEEEeqnarray}{cr}
  z_{dij} \ge y^{nm}_{dij}
  \ \ \ \
  \forall (i,j) \in A, \forall (n,m) \in F, \forall d \in D^{nm} &\label{eq:capuseBP}
\end{IEEEeqnarray}

These constraints are needed to evaluate the variable $z_{dij}$.

Having reviewed the main backup path formulation, we now present, in the the next subsection
a congestion avoidance formulation to be used for comparison purposes.

\subsection{Congestion Avoidance Formulation}
\label{sec:bp-ca-problem}

Let us first define the following additional variables:

\begin{description}[\IEEEsetlabelwidth{$D^{nm}_g$}]
  \item[$\mu_{ij}$] represents the maximum capacity used on link $(i,j)$ w.r.t. all possible failure detection events;
  \item[$\phi_{ij}$] represents the cost of using link $(i,j)$ when the capacity used is $\mu_{ij}$.
\end{description}

The problem can then be formulated as follows
  
\subsection*{Objective function}
  \begin{equation}
  min \sum_{(i,j) \in A} \phi_{ij} 
  \label{eq:ca-obj-func}
\end{equation}

This new objective function is a classical non-linear congestion related optimization function that aims at minimizing the load on each link.
As we will later see, the function will be linearized in order to treat the integer problem.

\subsection*{Link capacity constraints}

Previous constraints (\ref{eq:bp-la-cons-generic}), (\ref{eq:bp-la-cons-terminal}) and (\ref{eq:bp-fc-cons}) are maintained, while link capacity constrains (\ref{eq:bp-lc-cons-link}) and (\ref{eq:bp-lc-cons-node}) are substituted by the following:

\begin{IEEEeqnarray} {ccr}
  u^{nm}_{ij}
  + \sum_{d \in D^{nm}} b_d y^{nm}_{dij}
  + \sum_{e \in D^{mn}} b_e y^{mn}_{eij}    
  \le  \mu_{ij} \nonumber\\
  \forall (n,m) \in F, \forall (i,j) \in L
  \label{eq:ca-lc-cons-link}
\end{IEEEeqnarray}

\begin{IEEEeqnarray}{cr}
  v^{m}_{ij}
  + \sum_{\substack{n \in N: \\ (n,m) \in F}} \sum_{d \in D^{nm}} b_d y^{nm}_{dij} 
  \le \mu_{ij} \nonumber\\
  \forall m \in N, \forall (i,j) \in L
  \label{eq:ca-lc-cons-node}
\end{IEEEeqnarray}

\begin{equation} 
  \mu_{ij} \le w_{cap}c_{ij}
  \ \ \ \
  \forall (i,j) \in L
   \label{eq:ca-lc-cons}
\end{equation}

(\ref{eq:ca-lc-cons-link}) and (\ref{eq:ca-lc-cons-node}) evaluate the maximum load on link $(i,j)$ for all considered failure detection events $(m,n)$, while (\ref{eq:ca-lc-cons}) stipulates that even for the maximum value the capacity of the link must be respected.

\subsection*{Linearization constraints}

Given that $\phi_{ij}$ in (\ref{eq:ca-obj-func}) is a non-linear performance function, it should be linearized by the following constraints:

\begin{IEEEeqnarray} {clr}
  & \phi_{ij} \ge     \frac{\mu_{ij}}{w_{cap}c_{ij}}                & \ \ \ \ \forall (i,j) \in A \label{eq:cost1}\\
  & \phi_{ij} \ge 3   \frac{\mu_{ij}}{w_{cap}c_{ij}}-\frac{2}{3}    & \ \ \ \ \forall (i,j) \in A \label{eq:cost2} \\
  & \phi_{ij} \ge 10  \frac{\mu_{ij}}{w_{cap}c_{ij}}-\frac{16}{3}   & \ \ \ \ \forall (i,j) \in A \label{eq:cost3} \\
  & \phi_{ij} \ge 70  \frac{\mu_{ij}}{w_{cap}c_{ij}}-\frac{178}{3}  & \ \ \ \ \forall (i,j) \in A \label{eq:cost4} \\
  & \phi_{ij} \ge 500 \frac{\mu_{ij}}{w_{cap}c_{ij}}-\frac{1468}{3} & \ \ \ \ \forall (i,j) \in A \label{eq:cost5} 
\end{IEEEeqnarray}

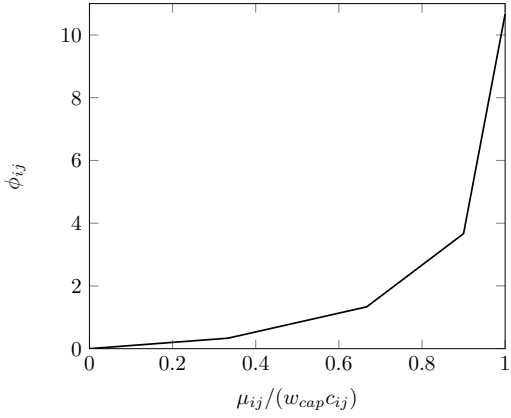
\begin{figure}
  \centering
  \resizebox{0.8\columnwidth}{!}{
    \begin{tikzpicture}
      \begin{axis}[ 
        xmin = -0, xmax = 1,
        ymin = 0, ymax = 11,
        xlabel=$\mu_{ij} / (w_{cap}c_{ij})$,
        ylabel=$\phi_{ij}$
      ]
      \addplot[black, thick]
      coordinates{
        (0,0)
        (0.333, 0.3333) 
        (0.667, 1.333)
        (0.9, 3.667)
        (1,10.667)
      };
      \end{axis}
    \end{tikzpicture}
  }
  \caption{Load cost function $\phi_{ij}$}
  \label{fig:load-cost-func}
  \vspace{-5mm}
\end{figure}

This set of equations represent the linearized load cost function shown in Fig. \ref{fig:load-cost-func}.

\section{Computational Results}
\label{sec:results}

\begin{table}
  \caption{Topologies summary}
  \centering
  \begin{tabular}{|c|c|c|c|c|c|}
    \hline
    \bfseries Topology &$\mid N \mid$ &$\mid A\mid$ & $\mid N_{edge}\mid$ &  $\mid N_{core}\mid$ &$\mid D\mid$ \\
    \hline
    Polska   & 12 & 36  & 9  & 3  & 72  \\
    Norway   & 27 & 102 & 16 & 11 & 240 \\
    Fat tree & 20 & 64  & 8  & 12 & 56  \\
    \hline
  \end{tabular}
  \label{table:topo-summary}
\end{table}

\begin{figure*}
  \centering
  \subfloat[][]{\includegraphics[width=0.27\textwidth]{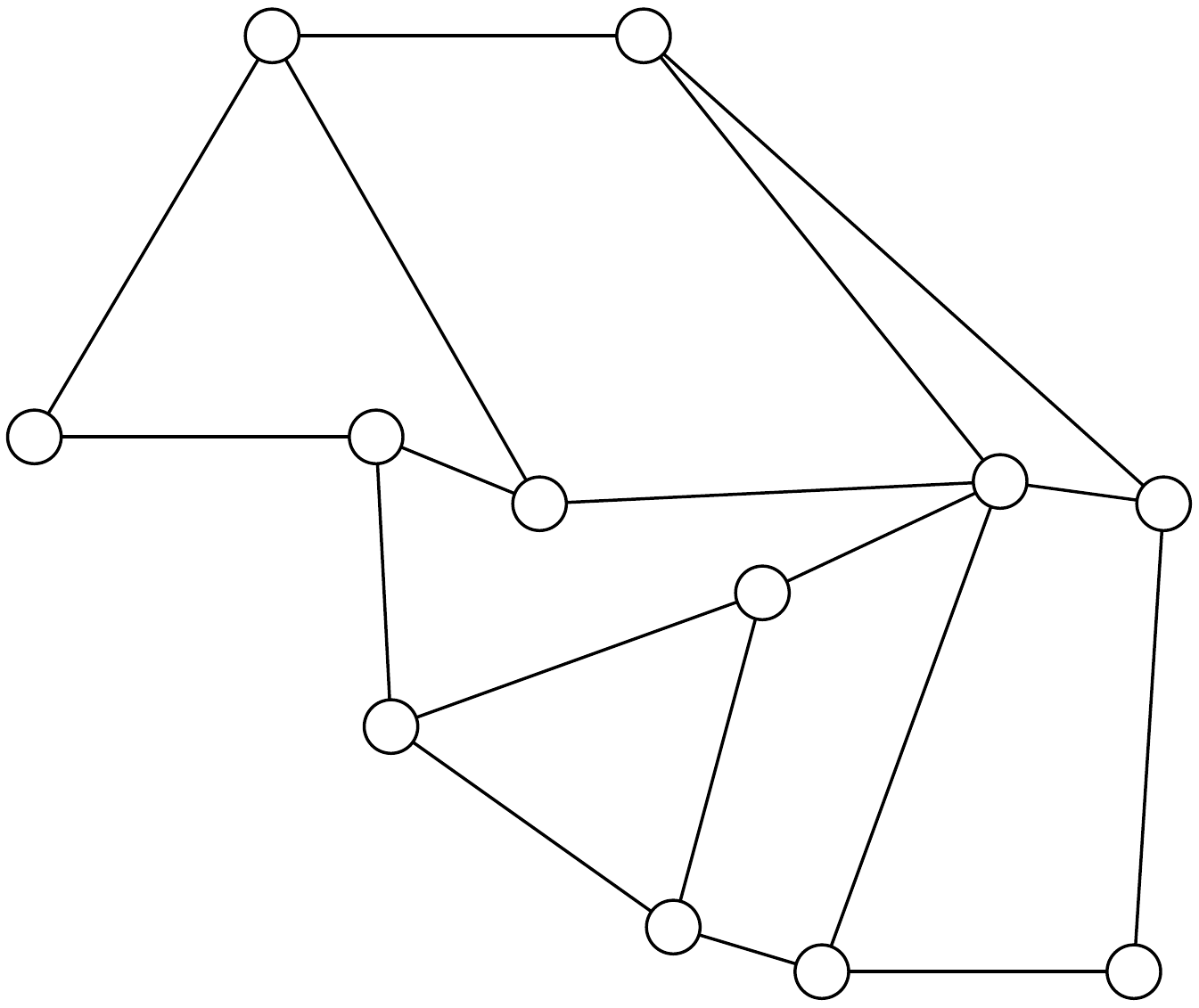}
    \label{fig:polska-top}}
  ~
  \subfloat[][]{\includegraphics[width=0.27\textwidth]{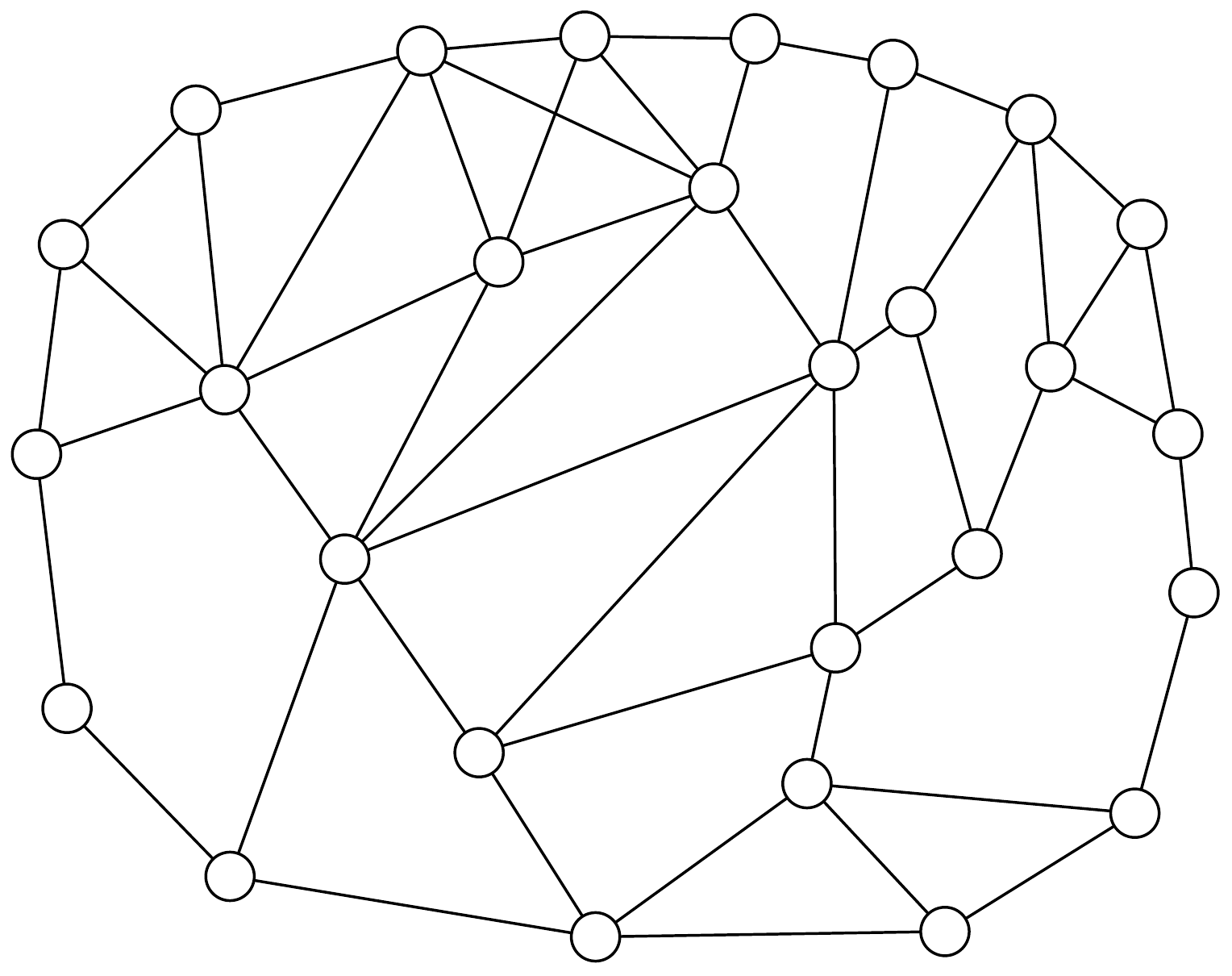}
    \label{fig:norway-top}}
  ~
  \subfloat[][]{\includegraphics[width=0.27\textwidth]{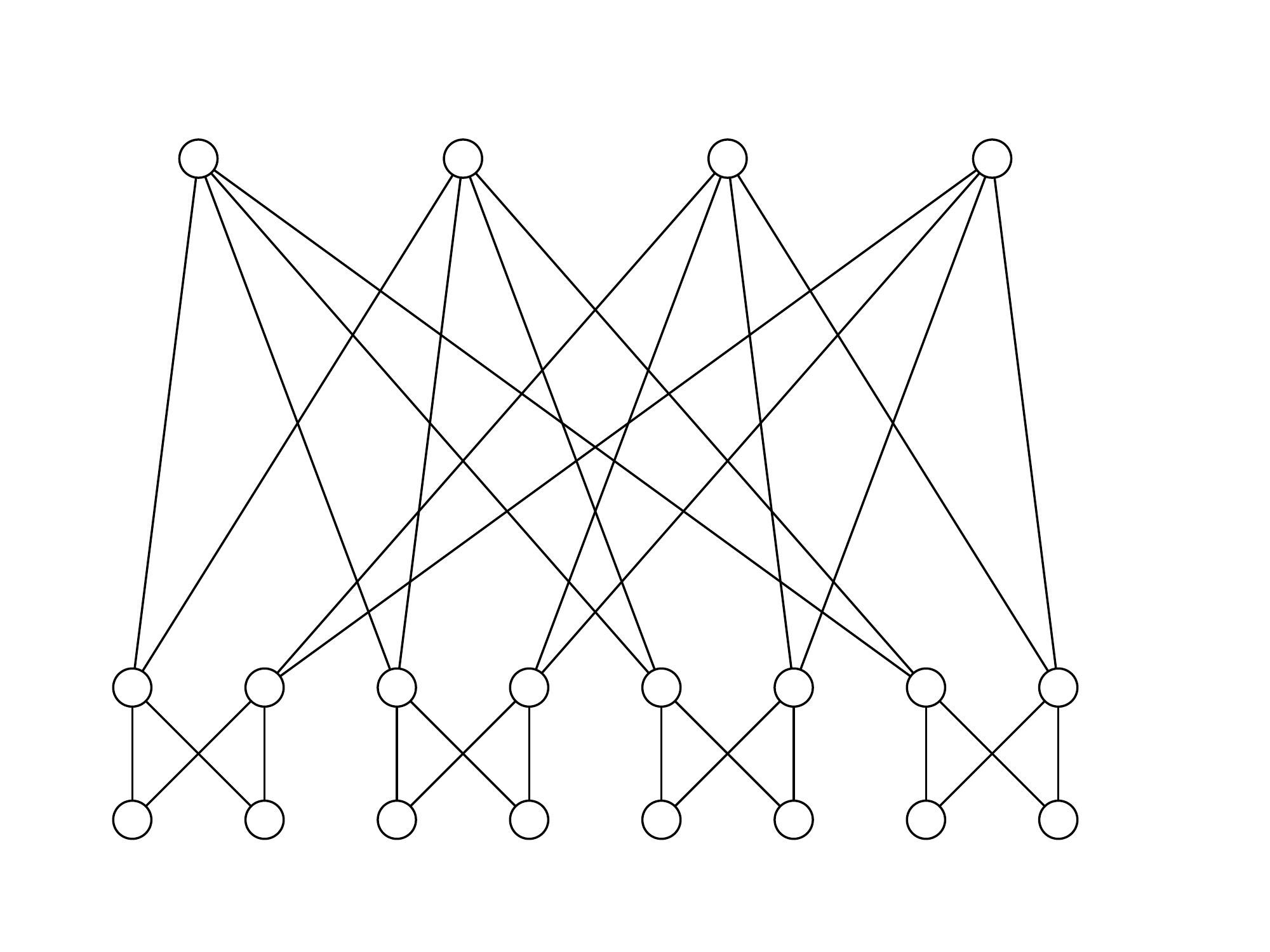}
    \label{fig:ft-top}}
  \caption{Network topologies used in test instances: (a) Polska, (b) Norway, and (c) Fat tree}%
  \label{fig:topologies}
  \vspace{-5mm}
\end{figure*}

The model was tested on three different network topologies portrayed in Figure \ref{fig:topologies}.  Two real backbone topologies, namely Polska and Norway, taken from \cite{orlowski10}, and a fat tree, which is an example of a symmetric topology well known for its degree of fault-resiliency \cite{niranjan09}, and widely used in data centers. For each topology, nodes are  divided in two sets: edge nodes and core nodes. Edge nodes act as source and destination of traffic while core nodes are only in charge of routing.

As mentioned in Section \ref{sec:model}, one of the inputs of the model is a set of primary paths evaluated as shortest paths for every traffic demand. Once such input was known, backup paths were found by varying weights $w_h$,$w_{y}$, and $w_{z}$ of objective function (\ref{eq:bp-obj}). Three types of instances were evaluated for comparison purposes: those referring to the backup problem with a given set of weights, those referring to the congestion avoidance formulation and those referring to a classic end-to-end path protection formulation. A summary of such instances is given below:

\begin{description}[\IEEEsetlabelwidth{BP$_{111}$}]
  \item[BP$_{111}$] all three terms of the objective function are taken into account;
  \item[BP$_{100}$] only the first term is considered, thus the model is forced to find a solution that minimizes the length of the reverse path, converging to a solution where failure detection node and reroute node are the same;
  \item[BP$_{010}$] only the second term is considered by minimizing the length of backup paths from $s_d$ to $t_d$;
  \item[BP$_{001}$] only the third term is considered, thus trying to minimize the number of links allocated for all backup paths of each demand;
  \item[BP$_{\text{CA}}$] congestion avoidance formulation of the BP problem, minimizing the maximum load for each link;
  \item[E2E] classic end-to-end path protection problem formulation.
\end{description}

The instances were executed assuming 2 different link capacity sets $c_{i,j}$: (i) capacity is set to the minimum value to obtain a feasible solution, and (ii) links are over-provisioned with very high capacity. For each test the requested bandwidth for each demand is always set to $b_d = 1$, and the available link capacity parameter is fixed to $w_{cap} = 80\%$.

The models were formalized and solved to optimality with AMPL-Cplex, using  PCs  with 8 CPU cores Intel Core i7 and 8GB of RAM. For all executions a solution was found in less than 30 seconds, except for the case of BP$_\text{CA}$ evaluated for the Norway topology, where the execution required about ten minutes.

The solutions were compared evaluating the trade-off with respect to the following parameters:

\begin{itemize}
  \item \textbf{Backup path length:} this measure was assessed with respect to the primary path length. A value of 100\% means that the length of the backup path is twice the primary path length, whereas 0\% indicates that the two paths have the same length.
  \item \textbf{Link capacity occupation:} is the percentage of the total link capacity allocated for all primary and backup paths that use the considered link.
  \item \textbf{Reverse path length:} is the portion of the primary path that a tagged packet has to traverse before being rerouted. A value of 100\%
  indicates that the packet has to go back to the source node of the demand, while a 0\% means that the packet is rerouted from the same node that detected the failure.
\end{itemize}

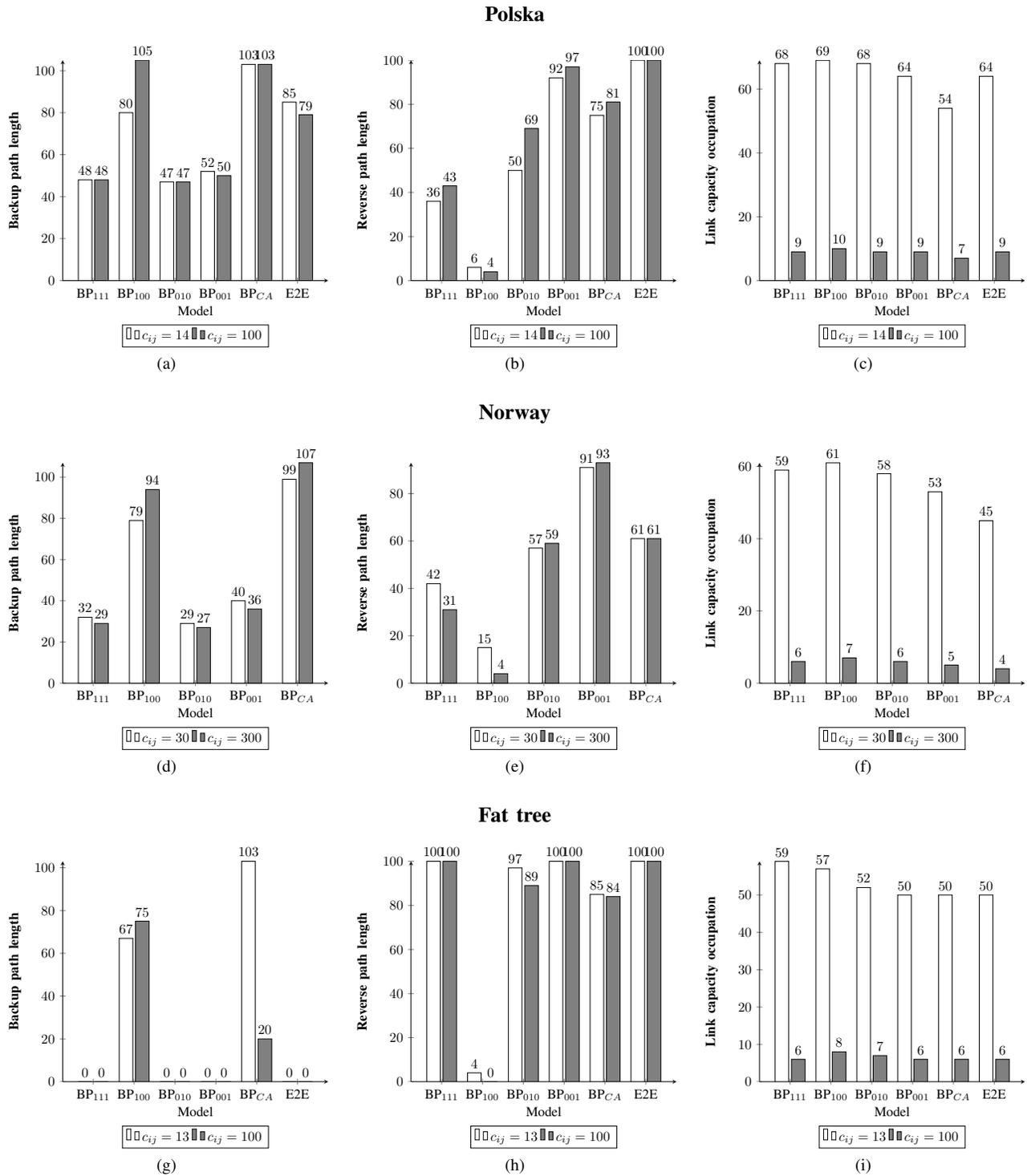
\begin{figure*}
   \centering
   \textbf{Polska}\\
   \subfloat[][]{
     \resizebox{0.3\textwidth}{!}{
       \begin{tikzpicture}
         \begin{axis}[
             ybar,
             axis x line=bottom,
             axis y line=left, ymin = 0,
             enlarge x limits=0.15,
             legend style={at={(0.5,-0.20)}, anchor=north,legend columns=-1},
            ylabel={\bfseries Backup path length},
             xlabel={Model},
             symbolic x coords={BP$_{111}$,BP$_{100}$,BP$_{010}$,BP$_{001}$,BP$_{CA}$,E2E},
             xtick=data,
             nodes near coords,
             nodes near coords align={vertical},
             ]
         \addplot[fill=white] coordinates {(BP$_{111}$,48) (BP$_{100}$,80) (BP$_{010}$,47) (BP$_{001}$,52) (BP$_{CA}$,103) (E2E,85)};
         \addplot[fill=gray] coordinates {(BP$_{111}$,48) (BP$_{100}$,105) (BP$_{010}$, 47) (BP$_{001}$,50) (BP$_{CA}$,103) (E2E,79)};
         \legend{$c_{ij} = 14$,$c_{ij} = 100$}
         \end{axis}
       \end{tikzpicture}
     }
     \label{fig:polska-pl}%
   }
   \subfloat[][]{
     \resizebox{0.3\textwidth}{!}{
       \begin{tikzpicture}
         \begin{axis}[
            ybar,
             axis x line=bottom,
             axis y line=left, ymin = 0,
             enlarge x limits=0.15,
             enlarge y limits=0,
             legend style={at={(0.5,-0.20)},anchor=north,legend columns=-1},
             ylabel={\bfseries Reverse path length},
             xlabel={Model},
             symbolic x coords={BP$_{111}$,BP$_{100}$,BP$_{010}$,BP$_{001}$,BP$_{CA}$,E2E},
             xtick=data,
             nodes near coords,
             nodes near coords align={vertical},
             ]
         \addplot [fill=white]coordinates {(BP$_{111}$,36) (BP$_{100}$,6) (BP$_{010}$,50) (BP$_{001}$,92) (BP$_{CA}$,75) (E2E,100)};
         \addplot [fill=gray] coordinates {(BP$_{111}$,43) (BP$_{100}$,4) (BP$_{010}$, 69) (BP$_{001}$,97) (BP$_{CA}$,81) (E2E,100)};
         \legend{$c_{ij} = 14$,$c_{ij} = 100$}
         \end{axis}
       \end{tikzpicture}
     }
     \label{fig:polska-rp}%
   }
  \subfloat[][]{
     \resizebox{0.3\textwidth}{!}{
       \begin{tikzpicture}
         \begin{axis}[
             ybar,
             axis x line=bottom,
             axis y line=left, ymin = 0,
             enlarge x limits=0.15,
             enlarge y limits=0,
             legend style={at={(0.5,-0.20)}, anchor=north,legend columns=-1},
            ylabel={\bfseries Link capacity occupation},
             xlabel={Model},
             symbolic x coords={BP$_{111}$,BP$_{100}$,BP$_{010}$,BP$_{001}$,BP$_{CA}$,E2E},
             xtick=data,
             nodes near coords,
             nodes near coords align={vertical},
             ]
         \addplot[fill=white] coordinates {(BP$_{111}$,68) (BP$_{100}$,69) (BP$_{010}$,68) (BP$_{001}$,64) (BP$_{CA}$,54) (E2E,64)};
         \addplot[fill=gray] coordinates {(BP$_{111}$,9) (BP$_{100}$,10) (BP$_{010}$, 9) (BP$_{001}$,9) (BP$_{CA}$,7) (E2E,9)};
         \legend{$c_{ij} = 14$,$c_{ij} = 100$}
         \end{axis}
       \end{tikzpicture}
     }
     \label{fig:polska-uc}%
   }
   \\
   \vspace{5mm}
   \textbf{Norway}\\
   \subfloat[][]{
     \resizebox{0.3\textwidth}{!}{
       \begin{tikzpicture}
         \begin{axis}[
             ybar,
             axis x line=bottom,
             axis y line=left, ymin = 0,
             enlarge x limits=0.15,
             legend style={at={(0.5,-0.20)}, anchor=north,legend columns=-1},
             ylabel={\bfseries Backup path length},
             xlabel={Model},
             symbolic x coords={BP$_{111}$,BP$_{100}$,BP$_{010}$,BP$_{001}$,BP$_{CA}$,E2E},
             xtick=data,
             nodes near coords,
             nodes near coords align={vertical},
             ]
         \addplot[fill=white] coordinates {(BP$_{111}$,32) (BP$_{100}$,79) (BP$_{010}$,29) (BP$_{001}$,40) (BP$_{CA}$,99)};
         \addplot[fill=gray] coordinates {(BP$_{111}$,29) (BP$_{100}$,94) (BP$_{010}$, 27) (BP$_{001}$,36) (BP$_{CA}$,107)};
         \legend{$c_{ij} = 30$,$c_{ij} = 300$}
         \end{axis}
       \end{tikzpicture}
     }
     \label{fig:norway-pl}%
   }
   \subfloat[][]{
     \resizebox{0.3\textwidth}{!}{
       \begin{tikzpicture}
         \begin{axis}[
            ybar,
             axis x line=bottom,
             axis y line=left, ymin = 0,
             enlarge x limits=0.15,
             enlarge y limits=0,
             legend style={at={(0.5,-0.20)},anchor=north,legend columns=-1},
             ylabel={\bfseries Reverse path length},
             xlabel={Model},
             symbolic x coords={BP$_{111}$,BP$_{100}$,BP$_{010}$,BP$_{001}$,BP$_{CA}$,E2E},
             xtick=data,
             nodes near coords,
             nodes near coords align={vertical},
             ]
         \addplot [fill=white]coordinates {(BP$_{111}$,42) (BP$_{100}$,15) (BP$_{010}$,57) (BP$_{001}$,91) (BP$_{CA}$,61)};
         \addplot[fill=gray] coordinates {(BP$_{111}$,31) (BP$_{100}$,4) (BP$_{010}$, 59) (BP$_{001}$,93) (BP$_{CA}$,61)};
         \legend{$c_{ij} = 30$,$c_{ij} = 300$}
         \end{axis}
       \end{tikzpicture}
     }
     \label{fig:norway-rp}%
   }
    \subfloat[][]{
     \resizebox{0.3\textwidth}{!}{
       \begin{tikzpicture}
         \begin{axis}[
             ybar,
             axis x line=bottom,
             axis y line=left, ymin = 0,
             enlarge x limits=0.15,
             enlarge y limits=0,
             legend style={at={(0.5,-0.20)}, anchor=north,legend columns=-1},
            ylabel={\bfseries Link capacity occupation},
             xlabel={Model},
             symbolic x coords={BP$_{111}$,BP$_{100}$,BP$_{010}$,BP$_{001}$,BP$_{CA}$,E2E},
             xtick=data,
             nodes near coords,
             nodes near coords align={vertical},
             ]
         \addplot[fill=white] coordinates {(BP$_{111}$,59) (BP$_{100}$,61) (BP$_{010}$,58) (BP$_{001}$,53) (BP$_{CA}$,45) };
         \addplot[fill=gray] coordinates {(BP$_{111}$,6) (BP$_{100}$,7) (BP$_{010}$, 6) (BP$_{001}$,5) (BP$_{CA}$,4)};
         \legend{$c_{ij} = 30$,$c_{ij} = 300$}
         \end{axis}
       \end{tikzpicture}
     }
     \label{fig:norway-uc}%
   }
   \\
   \vspace{5mm}
   \textbf{Fat tree}\\
   \subfloat[][]{
     \resizebox{0.3\textwidth}{!}{
       \begin{tikzpicture}
         \begin{axis}[
             ybar,
             axis x line=bottom,
             axis y line=left, ymin = 0,
             enlarge x limits=0.15,
             legend style={at={(0.5,-0.20)}, anchor=north,legend columns=-1},
            ylabel={\bfseries Backup path length},
             xlabel={Model},
             symbolic x coords={BP$_{111}$,BP$_{100}$,BP$_{010}$,BP$_{001}$,BP$_{CA}$,E2E},
             xtick=data,
             nodes near coords,
             nodes near coords align={vertical},
             ]
         \addplot[fill=white] coordinates {(BP$_{111}$,0) (BP$_{100}$,67) (BP$_{010}$,0) (BP$_{001}$,0) (BP$_{CA}$,103) (E2E,0)};
         \addplot[fill=gray] coordinates {(BP$_{111}$,0) (BP$_{100}$,75) (BP$_{010}$, 0) (BP$_{001}$,0) (BP$_{CA}$,20) (E2E,0)};
         \legend{$c_{ij} = 13$,$c_{ij} = 100$}
         \end{axis}
       \end{tikzpicture}
     }
     \label{fig:fattree-pl}%
   }
   \subfloat[][]{
     \resizebox{0.3\textwidth}{!}{
       \begin{tikzpicture}
         \begin{axis}[
            ybar,
             axis x line=bottom,
             axis y line=left, ymin = 0,
             enlarge x limits=0.15,
             enlarge y limits=0,
             legend style={at={(0.5,-0.20)},anchor=north,legend columns=-1},
             ylabel={\bfseries Reverse path length},
             xlabel={Model},
             symbolic x coords={BP$_{111}$,BP$_{100}$,BP$_{010}$,BP$_{001}$,BP$_{CA}$,E2E},
             xtick=data,
             nodes near coords,
             nodes near coords align={vertical},
             ]
         \addplot [fill=white]coordinates {(BP$_{111}$,100) (BP$_{100}$,4) (BP$_{010}$,97) (BP$_{001}$,100) (BP$_{CA}$,85) (E2E,100)};
         \addplot[fill=gray] coordinates {(BP$_{111}$,100) (BP$_{100}$,0) (BP$_{010}$, 89) (BP$_{001}$,100) (BP$_{CA}$,84) (E2E,100)};
         \legend{$c_{ij} = 13$,$c_{ij} = 100$}
         \end{axis}
       \end{tikzpicture}
     }
     \label{fig:fattree-rp}%
   }
    \subfloat[][]{
     \resizebox{0.3\textwidth}{!}{
       \begin{tikzpicture}
         \begin{axis}[
             ybar,
             axis x line=bottom,
             axis y line=left, ymin = 0,
             enlarge x limits=0.15,
             enlarge y limits=0,
             legend style={at={(0.5,-0.20)}, anchor=north,legend columns=-1},
             ylabel={\bfseries Link capacity occupation},
             xlabel={Model},
             symbolic x coords={BP$_{111}$,BP$_{100}$,BP$_{010}$,BP$_{001}$,BP$_{CA}$,E2E},
             xtick=data,
             nodes near coords,
             nodes near coords align={vertical},
             ]
         \addplot[fill=white] coordinates {(BP$_{111}$,59) (BP$_{100}$,57) (BP$_{010}$,52) (BP$_{001}$,50) (BP$_{CA}$,50) (E2E,50)};
         \addplot[fill=gray] coordinates {(BP$_{111}$,6) (BP$_{100}$,8) (BP$_{010}$, 7) (BP$_{001}$,6) (BP$_{CA}$,6) (E2E,6)};
         \legend{$c_{ij} = 13$,$c_{ij} = 100$}
         \end{axis}
       \end{tikzpicture}
     }
     \label{fig:fattree-uc}%
   }
   \caption{Result charts for the three topology examinated}%
   \label{fig:result-charts}%
   \vspace{-5mm}
 \end{figure*}

The complete set of results is shown in Table \ref{table:comp-results} and in chart form in Figure~\ref{fig:result-charts}.

In all instances BP$_{\text{111}}$ offers the best trade-off in terms of backup path length and reverse path length, with no major drawbacks. BP$_{\text{CA}}$ produces better solutions in terms of link capacity occupation, especially when considering instances with minimum capacity $c_{ij}$ (see Figures~\ref{fig:polska-uc}, \ref{fig:norway-uc} and \ref{fig:fattree-uc} for a clearer view).  The drawback of using BP$_{\text{CA}}$ is represented by longer backup paths. In fact, for Norway and Polska topologies, BP$_{\text{CA}}$ produces solutions with the longest backup paths, about the double in both cases (Figures~\ref{fig:polska-pl} and \ref{fig:norway-pl}). However, note that in the case of an on-line scenario BP$_{\text{CA}}$ would guarantee more residual capacity and thus a higher probability of accepting new traffic demands.

Concerning the reverse path length, the best solution is obtained with configuration BP$_{100}$ (Figures \ref{fig:polska-rp}, \ref{fig:norway-rp}, \ref{fig:fattree-rp}). The drawback in this case is represented by longer backup paths, about the double when compared to primary paths. It is interesting to note that for the fat tree topology with $c_{ij}=100$ (Figure~\ref{fig:fattree-rp}) BP$_{100}$ returns a solution with reverse path length equal to 0\%. This is worth mentioning because this solution would be suitable to be implemented with OpenFlow fast-failover, where detect node and reroute node are always the same. Unfortunately such a solution is not always feasible, as it strongly depends on topology and capacity constraints. Indeed, for all other cases, BP$_{100}$ is unable to provide a solution with 0\% reverse path length. Thanks to this result we can show how our solution based on OpenState, which is able to handle reverse paths, guarantees an higher degree of fault-resiliency when compared to a solution based on OpenFlow fast-failover.

It is also interesting to note that for the Norway topology the given set of primary paths has no feasible solution for the E2E model. This is due to the fact that in the classic formulation of E2E path protection, primary paths and backup paths must be evaluated at the same time, thus avoiding the situation where for a given primary path is impossible to find a completely disjoint backup path. We show in this case the flexibility of our approach by always providing a feasible solution.

Finally, it is interesting to note how for the case of the fat tree topology, the results obtained from BP$_{111}$ are the same of the E2E model, always having backup path length equal to primary paths, and reverse path length equal to 100\%. This means that in case of failure, packets will be always rerouted from the source node of the demand. In this case a solution adopting OpenState would guarantee less disruption thanks to the fact that nodes would be able to automatically switch to the backup path, whereas OpenFlow would require to forward packet to the controller to enable the backup path at the source node by installing the respective forwarding rules.

\begin{table*}[t]
  \centering
  \caption{Computational results}
  \begin{tabular}{|c|c|ccc|ccc|ccc|}
    \hline
    \bfseries Instance & \bfseries Model & \multicolumn{3}{c|}{\bfseries Backup path length} & \multicolumn{3}{c|}{\bfseries Link capacity occupation} & \multicolumn{3}{c|}{\bfseries Reverse path length } \\
    && \bfseries  \textit{min} & \textit{max} & \textit{avg (var)} & \textit{min} & \textit{max} & \textit{avg (var)} & \bfseries\textit{min} & \textit{max} & \textit{avg (var)} \\
    
    \hline\hline
    & BP$_{111}$ &0\% &300\%& 48\% (61\%)& 29\%&79\%&68\% (10\%)&0\%&100\%&36\% (41\%) \\
    & BP$_{100}$ & 0\% & 900\% & 80\% (103\%)&43\%&79\%&69\% (9\%)  & 0\% & 100\% &6\% (19\%)\\
    \bfseries Polska& BP$_{010}$ & 0\% & 300\% & 47\% (61\%)&43\%&79\%&68\% (9\%)  & 0\% & 100\% &50\% (45\%)\\
    \bfseries$ c_{ij} = 14, \forall (i,j) \in A$& BP$_{001}$ & 0\% & 300\% & 52\% (60\%)&43\%&79\%&64\% (12\%)  & 0\% & 100\% &92\% (24\%)\\
    & BP$_{\text{CA}}$ & 0\% & 700\% & 103\% (123\%)&7\%&79\%&54\% (20\%)  & 0\% & 100\% &75\% (43\%)\\
    & E2E & 0\% & 300\% & 85\% (75\%)&29\%&79\%&64\% (13\%)  & 100\% & 100\% &100\% (0\%)\\
    \hline
    & BP$_{111}$ & 0\% &300\%& 48\% (61\%)& 4\%&12\%&9\% (2\%)&0\%&100\%&43\% (45\%) \\
    & BP$_{100}$ & 0\% & 600\% & 105\% (118\%)&5\%&16\%&10\% (2\%)  & 0\% & 100\% &4\% (16\%)\\
    \bfseries Polska& BP$_{010}$ & 0\% & 300\% & 47\% (61\%)&6\%&12\%&9\% (1\%)  & 0\% & 100\% &69\% (43\%)\\
    \bfseries $ c_{ij} = 100, \forall (i,j) \in A$ & BP$_{001}$ & 0\% & 300\% & 50\% (61\%)&4\%&11\%&9\% (2\%)  & 0\% & 100\% &97\% (16\%)\\
    & BP$_{\text{CA}}$ & 0\% & 700\% & 103\% (136\%)&2\%&11\%&7\% (3\%)  & 0\% & 100\% &81\% (39\%)\\
    & E2E & 0\% & 300\% & 79\% (77\%)&3\%&12\%&9\% (2\%)  & 100\% & 100\% &100\% (0\%)\\
    \hline
    \hline
    & BP$_{111}$ & 0\% &500\%& 32\% (55\%)& 3\%&80\%&59\% (20\%)&0\%&100\%&42\% (43\%) \\
    & BP$_{100}$ & 0\% & 900\% & 79\% (98\%)&17\%&80\%&61\% (18\%)  & 0\% & 100\% &15\% (31\%)\\
    \bfseries Norway & BP$_{010}$ & 0\% & 500\% & 29\% (53\%)&7\%&80\%&58\% (20\%)  & 0\% & 100\% &57\% (42\%)\\
    \bfseries $ c_{ij} = 30, \forall (i,j) \in A$& BP$_{001}$ & 0\% & 500\% & 40\% (54\%)&7\%&80\%&53\% (20\%)  & 0\% & 100\% &91\% (25\%)\\
    & BP$_{\text{CA}}$ & 0\% & 1600\% & 99\% (137\%)&0\%&80\%&45\% (25\%)  & 0\% & 100\% &61\% (49\%)\\
    &E2E & - & - & - & - & - & - & - & - & - \\
    \hline
    & BP$_{111}$ & 0\% &500\%& 29\% (51\%)& 0\%&12\%&6\% (3\%)&0\%&100\%&31\% (39\%) \\
    & BP$_{100}$ & 0\% & 1400\% & 94\% (131\%)&1\%&14\%&7\% (3\%)  & 0\% & 100\% &4\% (17\%)\\
     \bfseries Norway & BP$_{010}$ & 0\% & 500\% & 27\% (52\%)&0\%&12\%&6\% (3\%)  & 0\% & 100\% &59\% (42\%)\\
    \bfseries $ c_{ij} = 300, \forall (i,j) \in A$& BP$_{001}$ & 0\% & 500\% & 36\% (53\%)&0\%&12\%&5\% (3\%)  & 0\% & 100\% &93\% (23\%)\\
    & BP$_{\text{CA}}$ & 0\% & 1400\% & 107\% (138\%)&1\%&10\%&4\% (3\%)  & 0\% & 100\% &61\% (49\%)\\
    &E2E & - & - & - & - & - & - & - & - & - \\
    \hline
    \hline
    & BP$_{111}$ & 0\% &0\%& 0\% (0\%)& 15\%&77\% &59\% (13\%)&100\%&100\%&100\% (0\%) \\
    & BP$_{100}$ & 0\% & 500\% & 67\% (70\%)&31\%&77\%&57\% (13\%)  & 0\% & 100\% &4\% (13\%)\\
    \bfseries Fat tree& BP$_{010}$ & 0\% & 0\% & 0\% (0\%)&23\%&77\%&52\% (13\%)  & 0\% & 100\% &97\% (18\%)\\
    \bfseries $ c_{ij} = 13, \forall (i,j) \in A$ & BP$_{001}$ & 0\% & 0\% & 0\% (0\%)&15\%&77\%&50\% (14\%)  & 0\% & 100\% &100\% (0\%)\\
    & BP$_{\text{CA}}$ & 0\% & 150\% & 103\% (128\%)&0\%&77\%&50\% (15\%)  & 0\% & 100\% &85\% (35\%)\\ 
    & E2E & 0\% & 0\% & 0\% (0\%)&15\%&77\%&50\% (15\%)  & 100\% & 100\% &100\% (0\%)\\
    \hline
    & BP$_{111}$ & 0\% &0\%& 0\% (0\%)& 1\%&11\%&6\% (2\%)&100\%&100\%&100\% (0\%) \\
    & BP$_{100}$ & 0\% & 400\% & 75\% (75\%)&3\%&12\%&8\% (2\%)  & 0\% & 0\% &0\% (0\%)\\
    \bfseries Fat tree & BP$_{010}$ & 0\% &0\% & 0\% (0\%)&2\%&12\%&7\% (2\%)  & 0\% & 100\% &89\% (31\%)\\
    \bfseries $ c_{ij} = 100, \forall (i,j) \in A$ & BP$_{001}$ & 0\% & 0\% & 0\% (0\%)&0\%&12\%&6\% (2\%)  & 100\% & 100\% &100\% (0\%)\\
    & BP$_{\text{CA}}$ & 0\% & 200\% & 20\% (35\%)&1\%&11\%&6\% (2\%)  & 0\% & 100\% &84\% (36\%)\\
    & E2E & 0\% & 0\% & 0\% (0\%)&0\%&12\%&6\% (3\%)  & 100\% & 100\% &100\% (0\%)\\
    \hline
  \end{tabular}
  \label{table:comp-results}
\end{table*}

\section{Conclusion}
\label{sec:conclusion}
In this paper we have presented a new failure management framework for SDN and a mathematical modeling approach specifically designed to exploit the capabilities of OpenState. The framework considers both single link and single node failure. The protection scheme is based on the idea that, upon failure detection, packets can be tagged and backtracked along the primary path to signal the failure to the first convenient reroute node, automatically establishing a detour path.
Such scheme aims at having zero packet loss after failure detection, and doesn't require controller intervention.
The models were tested on three well-known topologies and comparative results were obtained, showing the superiority of the scheme with respect to a classic end-to-end path protection scheme an with respect to an approach based on the OpenFlow fast-failover mechanism. We are currently working on the dimensioning problem and developing the OpenState application to experimentally validate the proposed solution.

\section*{Acknowledgment}
This work has been funded by NSERC Discovery Grant and by the European Community BEBA project. Luca Pollini and Davide Sanvito were part of the team that coded the algorithms in an OpenState emulator. We are grateful for their input that allowed us to assess upfront the feasibility of the proposed modeling approaches.

\bibliography{biblio}
\bibliographystyle{IEEEtran}

\end{document}